\documentclass[prd,aps,showpacs,onecolumn]{revtex4}
\usepackage{amssymb}
\usepackage{amsfonts}
\usepackage{amsmath}
\usepackage{graphicx}
\usepackage{bm}

\begin{document}

\title{Solvable rational extension of translationally shape invariant
potentials}
\author{Yves Grandati and Alain B\'{e}rard }
\affiliation{Institut de Physique, Equipe BioPhyStat, ICPMB, IF CNRS 2843, Universit\'{e}
Paul Verlaine-Metz, 1 Bd Arago, 57078 Metz, Cedex 3, France}

\begin{abstract}
\bigskip Combining recent results on rational solutions to the Riccati-Schr%
\"{o}dinger equations for shape invariant potentials to the scheme developed
by Fellows and Smith in the case of the one-dimensional harmonic oscillator,
we show that it is possible to generate an infinite set of solvable rational
extensions for every translationally shape invariant potential of second
category.
\end{abstract}

\maketitle

\bigskip

%\date{\today}

\section{Introduction}

In quantum mechanics there exists only few families of potentials which are
exactly solvable in closed-form. Most of them belong to the class of
shape-invariant potentials \cite{cooper,Dutt,Gendenshtein}. A possible way
to generate new solvable potentials is to start from the known ones and to
construct regular rational extensions of them. If the procedure has a long
history, in the last years important progress have been made in this
direction \cite{quesne,quesne2,odake}. A nice example of such a rational
extension is provided by the so-called CPRS potential \cite{carinena} which
is a rational extension of the one-dimensional harmonic oscillator. Very
recently Fellows and Smith \cite{fellows} showed that this potential can be
obtained as a supersymmetric partner of the harmonic oscillator. In the same
way they show how to generate an infinite family of partner potentials which
are regular rational extensions of the harmonic oscillator. This partnership
is based on the use of excited states Riccati-Schr\"{o}dinger (RS) functions
as superpotentials. This technique was devised for the first time by Robnik 
\cite{robnik,klippert}, but the potentials obtained are singular. Fellows
and Smith circumvent the problem by using a "spatial Wick rotation" which
eliminates the singularities from the real axis. In a recent work \cite%
{grandati} we propose a general scheme to obtain rational solutions to the
Riccati-Schr\"{o}dinger equations associated to the whole class of
translationally shape invariant potentials. These last are shared into two
categories which are related via simple changes of variables respectively to
the harmonic oscillator and to the isotonic oscillator. In this letter we
show how, by combining these results with the Robnik-Fellows-Smith
technique, we can generate an infinite set of regular rationally-extended
solvable potentials from every shape invariant potential of the second
category.

\section{Harmonic and isotonic oscillators}

\subsection{Basic scheme}

Let $H=-d^{2}/dx^{2}+V(x)$ of associated spectrum $\left( E_{n},w_{n}\right)
\equiv \left( E_{n},\psi _{n}\right) $, where $w_{n}(x)=-\psi _{n}^{\prime
}(x)/\psi _{n}(x)$.

The Riccati-Schr\"{o}dinger (RS) equation \cite{grandati} for the level $%
E_{n}$ is:

\begin{equation}
-w_{n}^{\prime }(x)+w_{n}^{2}(x)=V(x)-E_{n},  \label{edr4}
\end{equation}%
where we suppose $E_{0}=0$.

Make a "spatial Wick rotation", that is, set $x\rightarrow ix$, and define $%
v_{n}(x)=-iw_{n}(ix)$ Eq(\ref{edr4}) becomes:

\begin{equation}
v_{n}^{\prime }(x)+v_{n}^{2}(x)=V^{\left( n\right) }(x),
\end{equation}%
where:

\begin{equation}
V^{\left( n\right) }(x)=E_{n}-V(ix).
\end{equation}

$V^{\left( n\right) }(x)$ is supposed to be real and to have no movable
(that is $n$ dependent) singularity on the real line. Considering $v_{n}(x)$
as superpotential, $V^{\left( n\right) }$ can be viewed as the SUSY partner%
\cite{cooper} of $\widetilde{V}^{\left( n\right) }$ defined as:

\begin{equation}
\widetilde{V}^{\left( n\right) }(x)=V^{\left( n\right) }(x)-2v_{n}^{\prime
}(x)=2v_{n}^{2}(x)-V^{\left( n\right) }(x),
\end{equation}%
that is:

\begin{equation}
\widetilde{V}^{\left( n\right) }(x)=V(ix)-E_{n}+2v_{n}^{2}(x).
\end{equation}

The positive hamiltonians $H^{\left( n\right) }$ and $\widetilde{H}^{\left(
n\right) }$, associated to $V^{\left( n\right) }(x)$ and $\widetilde{V}%
^{\left( n\right) }(x)$ respectively, can be written:

\begin{equation}
\left\{ 
\begin{array}{c}
\widetilde{H}^{\left( n\right) }=A^{\left( n\right) +}A^{\left( n\right) }
\\ 
H^{\left( n\right) }=A^{\left( n\right) }A^{\left( n\right) +},%
\end{array}%
\right.
\end{equation}%
where:

\begin{equation}
A^{\left( n\right) }=\frac{d}{dx}+v_{n}(x).
\end{equation}

If $\widetilde{\psi }_{0}^{\left( n\right) }\left( x\right) \sim \exp \left(
-\int v_{n}(x)dx\right) $ is normalizable, it satisfies $A^{\left( n\right) }%
\widetilde{\psi }_{0}^{\left( n\right) }=0$ and is then the zero-energy
ground state of $\widetilde{H}^{\left( n\right) }$.

In this case the two hamiltonians are almost isospectral, that is:

\begin{equation}
\left\{ 
\begin{array}{c}
\widetilde{E}_{0}^{\left( n\right) }=0 \\ 
E_{k}^{\left( n\right) }=\widetilde{E}_{k+1}^{\left( n\right) },\ k\geq 0,%
\end{array}%
\right.
\end{equation}%
and their eigenstates are related by:

\begin{equation}
\left\{ 
\begin{array}{c}
\psi _{k}^{\left( n\right) }\left( x\right) =\frac{1}{\sqrt{\widetilde{E}%
_{k+1}^{\left( n\right) }}}A^{\left( n\right) }\widetilde{\psi }%
_{k+1}^{\left( n\right) }\left( x\right) \\ 
\widetilde{\psi }_{k}^{\left( n\right) }\left( x\right) =\frac{1}{\sqrt{%
E_{k}^{\left( n\right) }}}A^{\left( n\right) +}\psi _{k}^{\left( n\right)
}\left( x\right) .%
\end{array}%
\right.
\end{equation}

If $\widetilde{\psi }_{0}^{\left( n\right) }\left( x\right) $ is not
normalizable, the two hamiltonians are strictly isospectral, that is:

\begin{equation}
E_{k}^{\left( n\right) }=\widetilde{E}_{k}^{\left( n\right) },\ k\geq 0,
\end{equation}%
and their eigenstates are related by:

\begin{equation}
\left\{ 
\begin{array}{c}
\psi _{k}^{\left( n\right) }\left( x\right) =\frac{1}{\sqrt{E_{k}^{\left(
n\right) }}}A^{\left( n\right) }\widetilde{\psi }_{k}^{\left( n\right)
}\left( x\right) \\ 
\widetilde{\psi }_{k}^{\left( n\right) }\left( x\right) =\frac{1}{\sqrt{%
E_{k}^{\left( n\right) }}}A^{\left( n\right) +}\psi _{k}^{\left( n\right)
}\left( x\right)%
\end{array}%
\right. .
\end{equation}

Suppose that the potential considered satisfies the following identity (this
is the case of the harmonic and isotonic potentials):

\begin{equation}
-V(ix)=V(x)+\delta .
\end{equation}

We then have:

\begin{equation}
V^{\left( n\right) }(x)=V(x)+\delta +E_{n}.
\end{equation}

The spectrum of $H^{\left( n\right) }$ is:

\begin{equation}
\left\{ 
\begin{array}{c}
E_{k}^{\left( n\right) }=E_{k}+E_{n}+\delta \\ 
\psi _{k}^{\left( n\right) }\left( x\right) =\psi _{k}\left( x\right) \sim
\exp \left( -\int w_{k}(x)dx\right)%
\end{array}%
\right. ,\ k\geq 0.
\end{equation}

As for the spectrum of $\widetilde{H}^{\left( n\right) }$, it is either:%
\begin{equation}
\left\{ 
\begin{array}{c}
\widetilde{E}_{k}^{\left( n\right) }=E_{k}+E_{n}+\delta \\ 
\widetilde{\psi }_{k}^{\left( n\right) }\left( x\right) =\psi _{k}\left(
x\right)%
\end{array}%
\right. ,\ k\geq 0,
\end{equation}%
in the strictly isospectral case or:%
\begin{equation}
\left\{ 
\begin{array}{c}
\widetilde{E}_{0}^{\left( n\right) }=0 \\ 
\widetilde{E}_{k+1}^{\left( n\right) }=E_{k}+E_{n}+\delta%
\end{array}%
\right. ,\ k\geq 0,
\end{equation}%
with:

\begin{equation}
\left\{ 
\begin{array}{c}
\widetilde{\psi }_{0}^{\left( n\right) }\left( x\right) \sim \exp \left(
-\int v_{n}(x)dx\right) \\ 
\widetilde{\psi }_{k+1}^{\left( n\right) }\left( x\right) =\left(
E_{k}+E_{n}+\delta \right) ^{-1/2}A^{\left( n\right) +}\psi _{k}\left(
x\right)%
\end{array}%
\right. ,\ k\geq 0,
\end{equation}%
in the almost isospectral case.

We can illustrate this general scheme with two fundamental examples.

\subsection{Harmonic oscillator}

Consider the harmonic oscillator with zero ground-state energy:

\begin{equation}
V(x)=\frac{\omega ^{2}}{4}x^{2}-\frac{\omega }{2}.
\end{equation}

Its spectrum is well known:

\begin{equation}
\left\{ 
\begin{array}{c}
E_{n}=n\omega \\ 
\psi _{n}\left( x\right) \sim H_{n}\left( \omega x/2\right) \exp \left(
-\omega x^{2}/4\right)%
\end{array}%
\right.
\end{equation}%
and the corresponding RS functions $w_{n}(x)$ can be written as terminating
continued fractions \cite{grandati}. We then obtain for its "spatially Wick
rotated" image $v_{n}(x)=-iw_{n}(ix)$:

\begin{equation}
v_{n}(x)=\frac{\omega }{2}x+\frac{n\omega }{\omega x+}\Rsh ...\Rsh \frac{%
\left( n-j+1\right) \omega }{\omega x+}\Rsh ...\Rsh \frac{1}{x}.
\end{equation}

Clearly $v_{n}(x)$ does not present any singularity on the positive real
half line. The recurrence between the RS functions \cite{grandati} gives:

\begin{equation}
v_{n}(x)=v_{0}(x)+\frac{E_{n}}{v_{0}(x)+v_{n-1}(x)}
\end{equation}%
and $v_{n}(x)$ has the same odd parity as $v_{0}(x)$. Then $v_{n}(x)$ is
regular on all $\mathbb{R}$. Therefore the normalizability of $v_{n}(x)$ is
then ensured since the asymptotic behaviour at $\infty $ of $v_{n}(x)$ is
that of $v_{0}(x)$.

We have also:

\begin{equation}
-V(ix)=V(x)+\omega ,
\end{equation}%
that is, $\delta =\omega $ and:

\begin{equation}
V^{\left( n\right) }(x)=V(x)+\left( n+1\right) \omega .
\end{equation}

The spectrum of $H^{\left( n\right) }$ is then:

\begin{equation}
\left\{ 
\begin{array}{c}
E_{k}^{\left( n\right) }=\left( k+n+1\right) \omega \\ 
\psi _{k}^{\left( n\right) }\left( x\right) =\psi _{k}\left( x\right)%
\end{array}%
\right. ,\ k\geq 0.
\end{equation}

Its SUSY partner $\widetilde{H}^{\left( n\right) }$ has the following
associated potential:

\begin{equation}
\widetilde{V}^{\left( n\right) }(x)=2v_{n}^{2}(x)-\frac{\omega ^{2}}{4}%
x^{2}-\left( n+1\right) \omega
\end{equation}%
and constitutes a regular rational extension of $V(x)$ the spectrum of which
is completely determined. We have:

\begin{equation}
\left\{ 
\begin{array}{c}
\widetilde{E}_{0}^{\left( n\right) }=0 \\ 
\widetilde{E}_{k+1}^{\left( n\right) }=\left( k+n+1\right) \omega%
\end{array}%
\right. ,\ k\geq 0
\end{equation}%
and:

\begin{equation}
\left\{ 
\begin{array}{c}
\widetilde{\psi }_{0}^{\left( n\right) }\left( x\right) \sim \exp \left(
-\int v_{n}(x)dx\right) \\ 
\widetilde{\psi }_{k+1}^{\left( n\right) }\left( x\right) =\left( \left(
n+k+1\right) \omega \right) ^{-1/2}\left( -\frac{d}{dx}+v_{n}(x)\right) \psi
_{k}\left( x\right) .%
\end{array}%
\right.
\end{equation}

\subsection{Isotonic oscillator}

The potential of the isotonic oscillator with zero ground-state energy is:

\begin{equation}
V(x)=\frac{\omega ^{2}}{4}x^{2}+\frac{l(l+1)}{x^{2}}-\omega \left( l+\frac{3%
}{2}\right) ,\ x>0.
\end{equation}

Its spectrum is given by:

\begin{equation}
E_{n}=2n\omega ,\ \psi _{n}\left( x\right) \sim \exp \left( -\int
w_{n}\left( x\right) dx\right) ,
\end{equation}%
where the $w_{n}\left( x\right) $ are known \cite{grandati} and expressible
as terminating continued fractions. This gives:

\begin{equation}
v_{n}(x)=\frac{\omega }{2}x+\frac{l+1}{x}+\frac{2n\omega }{\omega x+\left(
2l+3\right) /x+}\Rsh ...\Rsh \frac{2\left( n-j+1\right) \omega }{\omega
x+\left( 2\left( l+j\right) +1\right) /x+}\Rsh ...\Rsh \frac{2\omega }{%
\omega x+\left( 2\left( l+n\right) -1\right) /x}.
\end{equation}

Clearly $v_{n}(x)$ does not present any singularity on the positive real
half line. It has to be noticed that, since

\begin{equation}
v_{0}(x)=\frac{\omega }{2}x+\frac{l+1}{x},  \label{fondIsot}
\end{equation}%
the term $\left( l+1\right) /x$ which then appears in every $v_{n}(x)$,
induces a nonnormalizable singularity at the origin for $\exp \left( -\int
v_{n}(x)dx\right) $. For instance:

\begin{equation}
\exp \left( -\int v_{0}(x)dx\right) =\frac{1}{x^{l+1}}\exp \left( -\frac{%
\omega }{4}x^{2}\right) .
\end{equation}

We are consequently in the case of a strict isospectrality.

We also have:

\begin{equation}
-V(ix)=V(x)+2\omega \left( l+\frac{3}{2}\right) ,
\end{equation}%
that is, $\delta =2\omega \left( l+\frac{3}{2}\right) $ and:

\begin{equation}
V^{\left( n\right) }(x)=\frac{\omega ^{2}}{4}x^{2}+\frac{l(l+1)}{x^{2}}%
+2\left( n+l+\frac{3}{2}\right) \omega .
\end{equation}

The spectrum of $H^{\left( n\right) }=-d^{2}/dx^{2}+V(x)+2\left( n+l+\frac{3%
}{2}\right) \omega $ is:

\begin{equation}
\left\{ 
\begin{array}{c}
E_{k}^{\left( n\right) }=2\left( k+n+l+\frac{3}{2}\right) \omega \\ 
\psi _{k}^{\left( n\right) }\left( x\right) =\psi _{k}\left( x\right) \sim
\exp \left( -\int w_{k}(x)dx\right)%
\end{array}%
\right. ,\ k\geq 0.
\end{equation}

Its SUSY partner $\widetilde{H}^{\left( n\right) }$ has the following
associated potential:

\begin{equation}
\widetilde{V}^{\left( n\right) }(x)=2v_{n}^{2}(x)-\frac{\omega ^{2}}{4}x^{2}-%
\frac{l(l+1)}{x^{2}}-2\left( n+l+\frac{3}{2}\right) \omega
\end{equation}%
and constitutes a regular rational extension of $V(x)$ the spectrum of which
is completely determined. We have:

\begin{equation}
\widetilde{E}_{k}^{\left( n\right) }=2\left( n+k+l+\frac{3}{2}\right) \omega
,\ k\geq 0\ 
\end{equation}%
and:

\begin{equation}
\widetilde{\psi }_{k}^{\left( n\right) }\left( x\right) =\frac{1}{\sqrt{%
2\left( n+k+l+\frac{3}{2}\right) \omega }}\left( -\frac{d}{dx}%
+v_{n}(x)\right) \psi _{k}\left( x\right) .  \notag
\end{equation}

\section{Second category potentials}

As shown in \cite{grandati}, the translationally shape invariant potentials
can be classified into two categories in which the potential can be brought
into a harmonic or isotonic form respectively, using a change of variables
which satisfy a constant coefficient Riccati equation. Consider the second
category. If we except the isotonic case itself, which has been treated
above, every potential of this category, with a zero ground-state energy $%
E_{0}=0$, is of the form \cite{grandati}:

\begin{equation}
V_{\pm }(y;a)=\lambda \left( \lambda \mp \alpha \right) y^{2}+\frac{\mu
\left( \mu -\alpha \right) }{y^{2}}+\lambda _{0\pm }(a)
\end{equation}%
with $a=\left( \lambda ,\mu \right) $, $\lambda _{0\pm }(a)=-\alpha \left(
\lambda \pm \mu \right) -2\lambda \mu $. The variable $y$ is defined via:

\begin{equation}
\frac{dy(x)}{dx}=\alpha \pm \alpha y^{2}(x),  \label{chvar}
\end{equation}%
that is, $y(x)=\tan \left( \alpha x+\varphi _{0}\right) ,$in the $V_{+}$
case ($+$ type) and $y(x)=\tanh \left( \alpha x+\varphi _{0}\right) $ or $%
y=\coth \left( \alpha x+\varphi _{0}\right) ,$ in the $V_{-}$ case ($-$
type).

The spectrum $\left( E_{n\pm },w_{\pm n}\right) $ of $H_{\pm
}=-d^{2}/dx^{2}+V_{\pm }(y;a)$ is known analytically \cite{grandati}. We
have for the energies:

\begin{equation*}
E_{n\pm }(a)=\pm \left( \phi _{2,\pm }\left( a_{n}\right) -\phi _{2,\pm
}\left( a\right) \right)
\end{equation*}%
with $\phi _{2,\pm }\left( a\right) =\left( \lambda \pm \mu \right) ^{2}$
and $a_{n}=\left( \lambda _{n},\mu _{n}\right) =\left( \lambda \pm n\alpha
,\mu +n\alpha \right) $.

As for the RS functions, they are given by:

\begin{eqnarray}
w_{n,\pm }(y,a) &=&\lambda y-\frac{\mu }{y}\mp \frac{\phi _{2,\pm }\left(
a_{n}\right) -\phi _{2,\pm }\left( a\right) }{\left( \lambda +\lambda
_{1}\right) y-\left( \mu +\mu _{1}\right) /y\mp }\Rsh ...
\label{RS functions cat 21} \\
&&  \notag \\
&\Rsh &\frac{\phi _{2,\pm }\left( a_{n}\right) -\phi _{2,\pm }\left(
a_{j-1}\right) }{\left( \lambda _{j-1}+\lambda _{j}\right) y-\left( \mu
_{j-1}+\mu _{j}\right) /y\mp }\Rsh ...  \notag \\
&&  \notag \\
&\Rsh &\frac{\phi _{2,\pm }\left( a_{n}\right) -\phi _{2,\pm }\left(
a_{n-1}\right) }{\left( \lambda _{n-1}+\lambda _{n}\right) y-\left( \mu
_{n-1}+\mu _{n}\right) /y}  \notag
\end{eqnarray}%
and in particular:

\begin{equation}
w_{0,\pm }(y;a)=\lambda y-\frac{\mu }{y}.
\end{equation}

The RS function $w_{\pm n}(y;a)$ associated the level $E_{\pm n}(a)$
satisfies:

\begin{equation}
-w_{\pm n}^{\prime }(x;a)+w_{\pm n}^{2}(x;a)=V_{\pm }(x;a)-E_{\pm n}(a)
\end{equation}%
or:

\begin{equation}
-\alpha \left( 1\pm y^{2}\right) w_{\pm n}^{\prime }(y;a)+w_{\pm
n}^{2}(y;a)=V_{\pm }(y,a)-E_{\pm n}(a).  \label{RSeq1}
\end{equation}

If we set $x\rightarrow ix$ and $y\rightarrow iy$, the change of variable Eq(%
\ref{chvar}) is transformed into $dy/dx=\alpha \mp \alpha y^{2}$.

Define $v_{\mp n}(y;a)=-iw_{\pm n}(iy;a)$ Eq(\ref{RSeq1}) becomes:

\begin{equation}
\alpha \left( 1\mp y^{2}\right) v_{\mp n}^{\prime }(y;a)+v_{\mp
n}^{2}(y;a)=V_{\mp }^{\left( n\right) }(y;a),  \label{RSeq2}
\end{equation}%
where:

\begin{eqnarray}
V_{\mp }^{\left( n\right) }(y;a) &=&E_{n\pm }(a)-V_{\pm }(iy;a) \\
&=&\lambda _{-1}\left( \lambda _{-1}\pm \alpha \right) y^{2}+\frac{\mu
\left( \mu -\alpha \right) }{y^{2}}+E_{n\pm }(a)-\lambda _{0\pm }(a)  \notag
\\
&=&V_{\mp }(y;\overline{a}_{\mp })+E_{n\pm }(a)-\left( \lambda _{0\pm
}(a)+\lambda _{0\mp }(\overline{a})\right)  \notag
\end{eqnarray}%
with $\lambda _{-1}=\lambda \mp \alpha $ and $\overline{a}=\left( \lambda
_{-1},\mu \right) $. We recover, up to a constant, a second category
potential but of the opposite of type and with a modified multiparameter.

The energy spectrum of $H_{\pm }^{\left( n\right) }=-d^{2}/dx^{2}+V_{\pm
}^{\left( n\right) }(y;a)$ is:

\begin{equation}
\left\{ 
\begin{array}{c}
E_{k\pm }^{\left( n\right) }=E_{n\pm }(a)+E_{k\mp }(\overline{a})-\left(
\lambda _{0\pm }(a)+\lambda _{0\mp }(\overline{a})\right) \\ 
\psi _{k\pm }^{\left( n\right) }\left( x\right) =\psi _{k\mp }\left( x\right)%
\end{array}%
\right. ,\ k\geq 0.
\end{equation}

Eq(\ref{RSeq2}) can be written as:

\begin{equation}
v_{\pm n}^{\prime }(x;a)+v_{\pm n}^{2}(x;a)=V_{\pm }^{\left( n\right) }(x;a)
\end{equation}%
and $v_{\pm n}(x;a)$ is the superpotential associated with $V_{\pm }^{\left(
n\right) }$. $H_{\pm }^{\left( n\right) }$is therefore the SUSY partner of $%
\widetilde{H}_{\pm }^{\left( n\right) }$ given by:\bigskip 
\begin{equation*}
\widetilde{H}_{\pm }^{\left( n\right) }=-\frac{d^{2}}{dx^{2}}+\widetilde{V}%
_{\pm }^{\left( n\right) }(x;a),
\end{equation*}%
where:\bigskip 
\begin{eqnarray}
\widetilde{V}_{\pm }^{\left( n\right) }(x;a) &=&V_{\pm }^{\left( n\right)
}(x;a)-2v_{\pm n}^{\prime }(x;a) \\
&=&2v_{\pm n}^{2}(x;a)-V_{\pm }^{\left( n\right) }(x;a),  \notag
\end{eqnarray}%
that is:

\begin{equation}
\widetilde{V}_{\pm }^{\left( n\right) }(x;a)=2v_{\pm n}^{2}(x;a)-V_{\mp }(y;%
\overline{a})-E_{n\pm }(a)+\left( \lambda _{0\pm }(a)+\lambda _{0\mp }(%
\overline{a})\right) .
\end{equation}

\bigskip The two hamiltonians $H_{\pm }^{\left( n\right) }$ and $\widetilde{H%
}_{\pm }^{\left( n\right) }$ are factorizable as:

\begin{equation}
\left\{ 
\begin{array}{c}
\widetilde{H}_{\pm }^{\left( n\right) }=A_{\pm }^{\left( n\right) +}A_{\pm
}^{\left( n\right) } \\ 
H_{\pm }^{\left( n\right) }=A_{\pm }^{\left( n\right) }A_{\pm }^{\left(
n\right) +},%
\end{array}%
\right.
\end{equation}%
where $A_{\pm }^{\left( n\right) }=d/dx+v_{\pm n}(x;a)=\left( \alpha \mp
\alpha y^{2}\right) d/dy+v_{\pm n}(y;a)$.

Both are positive definite and isospectral. They are even strictly
isospectral as in the case of the isotonic oscillator. Indeed we have:

\begin{equation}
v_{\pm 0}(y;a)=\lambda y+\frac{\mu }{y},
\end{equation}%
that is, $\exp \left( \int v_{\pm 0}(x;a)dx\right) $ and $\exp \left( \int
v_{\pm n}(x;a)dx\right) $ present a nonnormalizable singularity at the zero
of $y(x)$.

Consequently the energy spectrum of $\widetilde{H}_{\pm }^{\left( n\right) }$
is:

\begin{equation}
\widetilde{E}_{\pm k}^{\left( n\right) }=E_{\pm k}^{\left( n\right)
}=E_{n\pm }(a)+E_{k\mp }(\overline{a})-\left( \lambda _{0\pm }(a)+\lambda
_{0\mp }(\overline{a})\right) ,\ k\geq 0,  \label{ener2}
\end{equation}%
and the corresponding eigenstates are given by:

\begin{equation}
\widetilde{\psi }_{\pm k}^{\left( n\right) }\left( x\right) =\frac{1}{\sqrt{%
E_{\pm k}^{\left( n\right) }}}A_{\pm }^{\left( n\right) +}\psi _{k\pm
}^{\left( n\right) }\left( x\right) =\frac{1}{\sqrt{E_{\pm k}^{\left(
n\right) }}}A_{\pm }^{\left( n\right) +}\psi _{k\mp }\left( x\right) .
\label{fctpropres2}
\end{equation}

\bigskip Then for every $n$ the potential:%
\begin{equation}
\widetilde{V}_{\pm }^{\left( n\right) }(x;a)=2v_{\pm n}^{2}(x;a)-V_{\mp }(x;%
\overline{a})-E_{n\pm }(a)+\left( \lambda _{0\pm }(a)+\lambda _{0\mp }(%
\overline{a})\right) ,
\end{equation}%
where:

\begin{eqnarray}
v_{n,\pm }(y,a) &=&\lambda y+\frac{\mu }{y}\pm \frac{\phi _{2,\pm }\left(
a_{n}\right) -\phi _{2,\pm }\left( a\right) }{\left( \lambda +\lambda
_{1}\right) y+\left( \mu +\mu _{1}\right) /y\pm }\Rsh ... \\
&&  \notag \\
&\Rsh &\frac{\phi _{2,\pm }\left( a_{n}\right) -\phi _{2,\pm }\left(
a_{j-1}\right) }{\left( \lambda _{j-1}+\lambda _{j}\right) y+\left( \mu
_{j-1}+\mu _{j}\right) /y\pm }\Rsh ...  \notag \\
&&  \notag \\
&\Rsh &\frac{\phi _{2,\pm }\left( a_{n}\right) -\phi _{2,\pm }\left(
a_{n-1}\right) }{\left( \lambda _{n-1}+\lambda _{n}\right) y+\left( \mu
_{n-1}+\mu _{n}\right) /y},  \notag
\end{eqnarray}%
constitutes a solvable regular rational extension of $V_{\pm }(x;a)$.

\section{Conclusion}

We have shown that every translationally shape invariant potential of second
category admits an infinite family of solvable regular rational extensions.
All the members of this family are strictly isospectral to the original
potential and the associated eigenstates are easily related to the initial
ones by application of first order differential operators. The adaptation of
the above scheme of extension to the case of shape invariant potentials of
the first category is in progress.

\section{Acknowledgments}

We would like to thank Professor P.G.L. Leach for useful suggestions and a
careful reading of the manuscript.

\end{document}